\documentclass[prd,amsmath,amsfonts,showpacs,preprintnumbers,floatfix,twocolumn,a4paper]{revtex4}
\usepackage{graphicx}
\newcommand{\be}{\begin{eqnarray}}
\newcommand{\ee}{\end{eqnarray}}
\newcommand{\bi}{\begin{itemize}} 
\newcommand{\ei}{\end{itemize}}

\begin{document}

%
\preprint{ADP-11-41/T763}

\title{SU(3) centre vortices underpin confinement and dynamical chiral symmetry breaking}


\author{Elyse-Ann O'Malley}
\author{Waseem Kamleh}
\author{Derek Leinweber}
\author{Peter Moran}
\affiliation{Centre for the Subatomic Structure of Matter 
  (CSSM), School of Chemistry \& Physics, University of Adelaide 5005,
  Australia} 


\begin{abstract}
The mass function of the nonperturbative quark propagator in $SU(3)$
gauge theory shows only a weak dependence on the vortex content of the
gauge configurations.  Of particular note is the survival of dynamical
mass generation on vortex-free configurations having a vanishing
string tension.  This admits the possibility that mass generation
associated with dynamical chiral symmetry breaking persists without
confinement.  In this paper we examine the low-lying ground state
hadron spectrum of the $\pi$, $\rho$, $N$ and $\Delta$ and discover
that while dynamical mass generation persists in the vortex-free
theory, it is not connected to dynamical chiral symmetry breaking.  In
this way, centre vortices in $SU(3)$ gauge theory are intimately
linked to both confinement and dynamical chiral symmetry breaking.  
\end{abstract}



\maketitle


\section{Introduction}

Numerical simulations of QCD on a space-time lattice reveal that the
essential, fundamentally important, non-perturbative features of the
QCD vacuum fields are:
\begin{enumerate}
\item The dynamical generation of mass through chiral symmetry
  breaking ($\chi$SB), and 
\item The confinement of quarks.
\end{enumerate}
However, there exists no derivation of quark confinement starting from
first principles, nor is there a totally convincing explanation of the
effect.

The questions that dominate the field centre around gaining an
understanding on how these fundamentally important features of QCD
come about.  The question is: {\em What is the essence of QCD vacuum
  structure?}
That is, what is it about the field fluctuations of the QCD vacuum
that causes quarks to be confined?  What aspects of the QCD vacuum are
responsible for dynamical mass generation?  Do the underlying
mechanisms share a common origin?

The prevailing view is that quark confinement and dynamical $\chi$SB
is the work of some special class of gauge field configurations which
dominate the QCD vacuum on large distance scales.  Candidates have
included instantons, Abelian monopoles, and centre vortices.
In recent years, algorithms have been invented which can locate these
types of objects in thermalized lattices, generated by the lattice
Monte Carlo technique.  This is an important development enabling {\it
ab initio} investigations of the underlying mechanism of quark
confinement and dynamical $\chi$SB.

Instantons are natural candidates to explain $\chi$SB as each
instanton is associated with a zero mode of the Dirac operator
\cite{'tHooft:1976fv}.  An accumulation of zero eigenvalues will produce a
quark condensate \cite{Banks:1979yr}.  However, instantons are no
longer favored to play a significant role in confinement
\cite{Chen:1998ct}.

An attractive mechanism for confinement is dual superconductivity of
the QCD vacuum \cite{'tHooft:1981ht,Mandelstam:1974pi}.  
The condensation of chromo-magnetic monopoles has been observed
directly after gauge-fixing to ``Maximal Abelian Gauge''
\cite{Shiba:1994ab}, from which the idea of ``Abelian dominance'' has
emerged.  There, the Abelian degrees of freedom of the Yang-Mills
field are thought to encode all its long-distance properties.
However, degrees of freedom more elementary than Abelian monopoles,
embedded in them and solely responsible for the physics assigned to
them, cannot be ruled out \cite{deForcrand:1999ms}.

Like Abelian dominance, centre vortices are exposed by gauge-fixing.
Usually, a gauge transformation is applied which brings each lattice
link as close as possible to a centre element of the gauge group.  The
centre of a group is that set of group elements which commute
with all other elements of the group.  For an $SU(N)$ gauge group, the
centre elements consist of all $g\in SU(N)$ proportional to the
$N\times N$ unit matrix, $\mathbf{I}$, subject to the condition that
$\det(g)=1$.  This is the set of $N$ $SU(N)$ group elements $\{Z_m\},$ with
\be
       Z_m  = \exp \left(i \frac{2\pi}{N} m \right) \,
              \mathbf{I},
       ~~~~ (m=0,1,2,...,N-1) \, .
\ee
These centre elements form a discrete Abelian subgroup known as $Z_N$.
Vortices are identified as the defects in the centre-projected gauge
field.  Again, the idea of centre dominance is that the centre degree
of freedom encodes all the long-distance nonperturbative physics.

In $SU(2)$ gauge theory, a clear link between centre vortices,
confinement and mass generation via dynamical chiral symmetry breaking
is manifest \cite{Bowman:2008qd}.  Centre vortices are the single
underlying mechanism giving rise to both chiral symmetry breaking and
quark confinement in $SU(2)$ gauge theory.

Whether this is the case for the $SU(3)$ Yang-Mills theory relevant to QCD
is not as clear.  As outlined in
Refs.~\cite{Leinweber:2006zq,Cais:2007bm,Bowman:2010zr} the relation
between centre vortices and dynamical chiral symmetry breaking is much
more complicated in $SU(3)$ gauge theory.  Ref.~\cite{Bowman:2010zr}
explores the role of centre vortices identified by gauge fixing Monte
Carlo generated configurations to maximal centre gauge
\cite{DelDebbio:1998uu}, clearly illustrating how dynamical mass
generation survives the removal of these vortices.  This admits the
possibility that the underlying mechanisms generating confinement and
dynamical chiral symmetry breaking are decoupled.

We proceed to investigate the low-lying hadron mass spectrum in
this unique centre-vortex-free environment lacking confinement and
retaining dynamical mass generation.  A brief report on this was
presented at Lattice 2011 \cite{OMalleyLatt2011}.  Our aim here is
to search for evidence of dynamical chiral symmetry breaking and thus
provide further insight into the role of centre vortices in QCD.

\section{Centre Vortices}

There are multiple methods of identifying centre vortices, such as the
Maximal Centre
Gauge~\cite{DelDebbio:1996mh,Langfeld:1997jx,Langfeld:2003ev} and
Laplacian Centre Gauge \cite{deForcrand:2000pg} with various
preconditioning options \cite{Cais:2008za}.  Here we focus on vortices
identified by gauge fixing the original Monte-Carlo generated
configurations directly to Maximal Centre Gauge without any
preconditioning. This is the same identification used in
Ref.~\cite{Bowman:2010zr}, which revealed the survival of dynamical
mass generation on such vortex-free configurations.

First the links $U_\mu(x)$ are gauge transformed to be brought close
to the centre elements of $SU(3)$,
\be 
Z = \exp \left ( 2 \pi i\, \frac{m}{3} \right ) \, \mathbf{I}, \textrm{ with } m = -1, 0, 1.
\label{CentreSU3}
\ee
On the lattice this is implemented by searching for the gauge
transformation $\Omega$ such that,
\be
\sum_{x,\mu} \,  \left | \mathrm{tr}\, U_\mu^\Omega(x) \, \right |^2 \stackrel{\Omega}{\to}
\mathrm{max} \, .
\label{GaugeTrans}
\ee
One can then project the gluon field to a centre-vortex only
configuration where each link is a number (one of the roots of unity)
times the identity matrix
\be 
U_\mu(x) \to Z_\mu(x) \textrm{ where }
Z_\mu(x) = \exp \left ( 2 \pi i\, \frac{m_\mu(x)}{3} \right
)\mathbf{I} \, ,
\label{UtoZ}
\ee
where $m_\mu(x) = -1, 0, 1.$

The vortices are identified by the centre charge, $z$, found
by taking the product of the links around a plaquette,
\be
z = \prod_\Box Z_\mu(x) = \exp \left ( 2 \pi i\, \frac{n}{3} \right ) \, .
\label{CentreCharge}
\ee
If $z=1$, no vortex pierces the plaquette. If $z \neq 1$ a vortex with
charge $z$ pierces the plaquette.  In the smooth gauge-field limit,
all the links approach the identity, and no vortices are found.  It is
only when we get a non-trivial change of phase around the plaquette
that a vortex is identified.

Vortices are removed by removing the centre phase.  This is
done by making the transformation 
\be
U_\mu(x) \to U_\mu^\prime(x) = Z_\mu^*(x)\, U_\mu(x) \, .
\label{RemoveCV}
\ee 

In $SU(2)$ gauge theory the removal of the centre vortices results in
a lack of string tension which is fully recovered in the vortex-only
configurations.  The mass function of the nonperturbative quark
propagator observed in the vortex-removed configurations is flat and
shows no sign of dynamical mass generation.

The findings in $SU(3)$ gauge theory\cite{Bowman:2010zr} differ from
the $SU(2)$ case. While the removal of centre vortices removes the
string tension, the string tension is not fully recovered in the
vortex-only configurations.  An examination of the mass function of
the nonperturbative quark propagator reveals only small differences in
dynamical mass generation between the original and vortex-removed
configurations.  This shape indicates the retention of dynamical mass
generation, despite the absence of confinement.  This leads to the key
question under investigation.  Is the persistence of dynamical mass
generation a manifestation of dynamical chiral symmetry breaking in
the absence of confinement?

We note that at large momenta, the mass function of the propagator of the
vortex-removed configurations experiences a vertical shift upwards of
approximately 60 MeV for a given bare quark mass.  This may be
attributed to a roughening of the configurations at short distances
associated with the removal of centre vortices via Eq.~\ref{RemoveCV}.

\section{Results}

To further investigate the underlying physics, we calculate the
standard effective masses for low-lying hadrons from their Euclidean
two-point functions.  We compare the effective masses for the $\pi$,
$\rho$, $N$ and $\Delta$ hadrons obtained from the regular and
vortex-removed configurations for varying quark masses.  Our
uncertainties are obtained via the jackknife analysis with best fits
obtained through a consideration of the full covariance-matrix based
$\chi^2$ per degree of freedom.

A statistical ensemble of 200 $SU(3)$ gauge-field configurations is
generated using the L{\"u}scher-Weisz \cite{Luscher:1984xn} mean-field
improved action on a $20^3 \times 40$ lattice with a lattice spacing
of 0.125 fm.  We use the FLIC fermion action \cite{Zanotti:2001yb}
providing nonperturbative ${\mathcal O}(a)$ improvement
\cite{Zanotti:2004dr} with improved chiral properties allowing
efficient access to the light quark-mass regime
\cite{Boinepalli:2004fz}.

Initially we consider four different values for the Wilson hopping
parameter, $\kappa$, selected to provide a wide view of the mass
dependence of the spectrum.  We consider $\kappa = 0.1280$, 0.1293,
0.1304 and 0.1310.  The associated quark mass can be estimated by
linearly extrapolating the squared pion mass to zero as a function
$1/\kappa.$ The critical hopping parameter $\kappa_{\rm
  cr}$ is the value at which pion mass vanishes, such that
\be
m_q = \frac{1}{2 a} \left ( \frac{1}{\kappa} - \frac{1}{\kappa_{\rm
      cr}} \right ) \, .
\ee

\begin{figure}[tb]
\centering

\includegraphics[height = 0.48\hsize,angle=90]{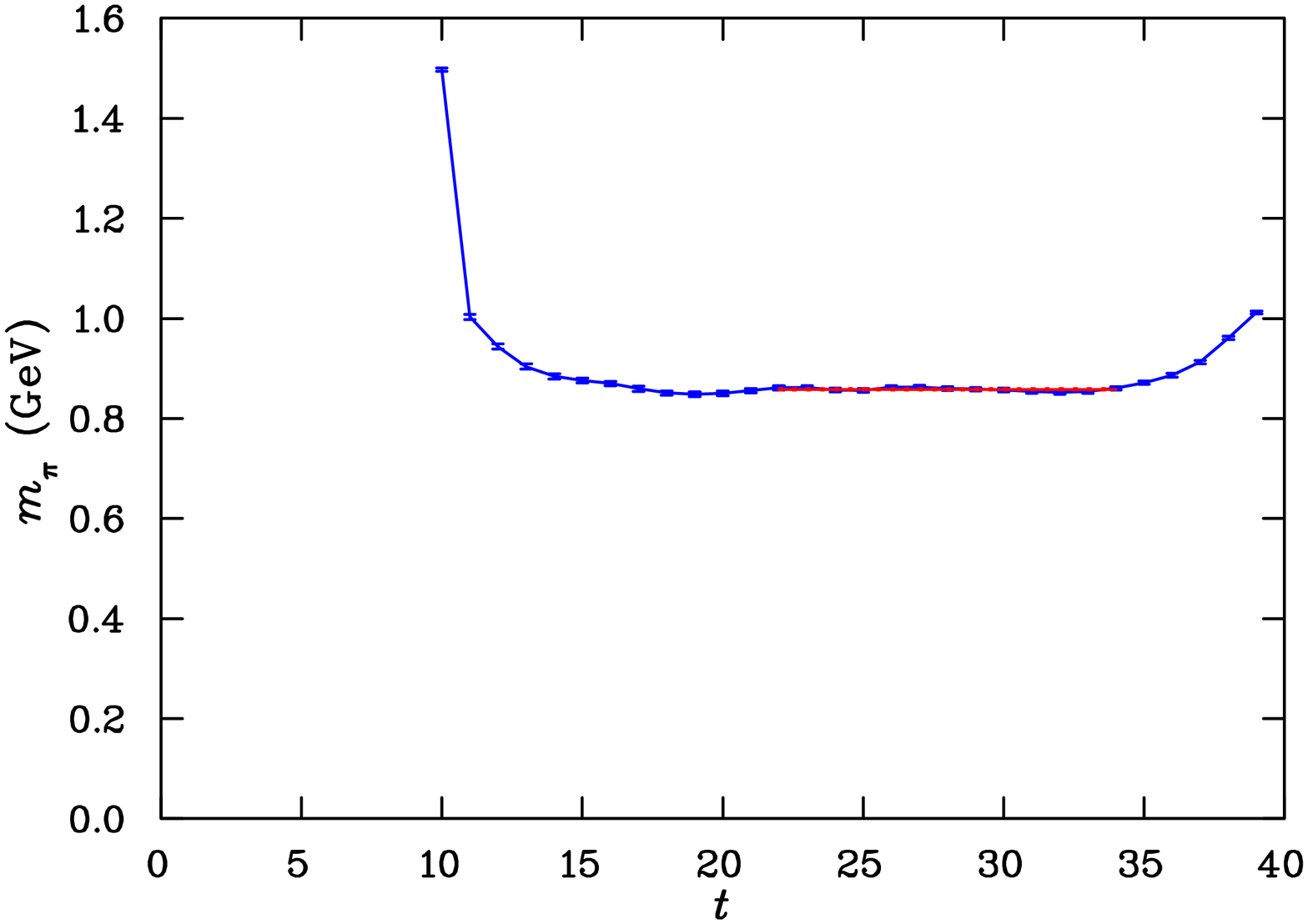} 
\hspace{0.09mm}
\includegraphics[height = 0.48\hsize,angle=90]{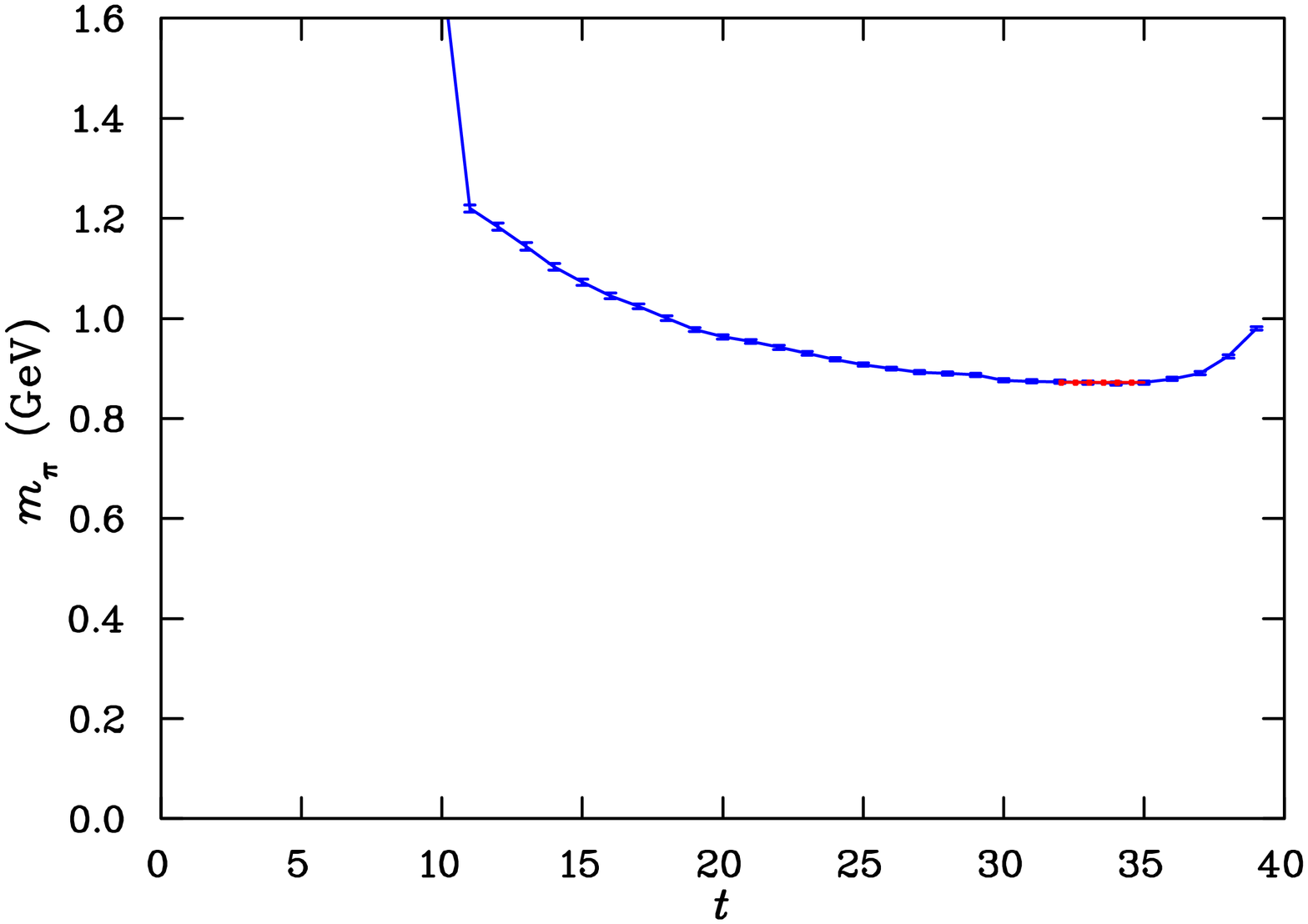}

\includegraphics[height = 0.48\hsize,angle=90]{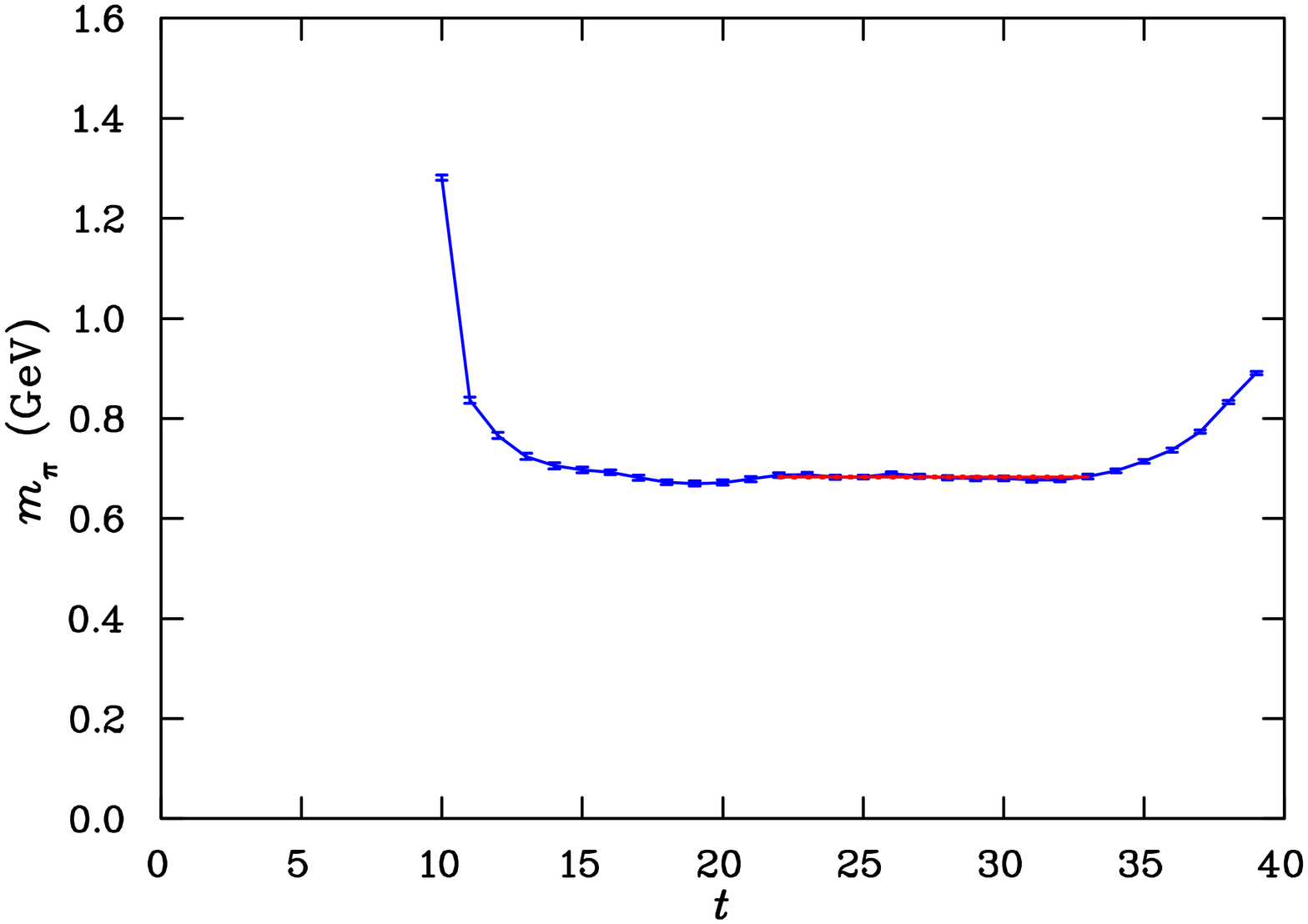}
\hspace{0.09mm}
\includegraphics[height = 0.48\hsize,angle=90]{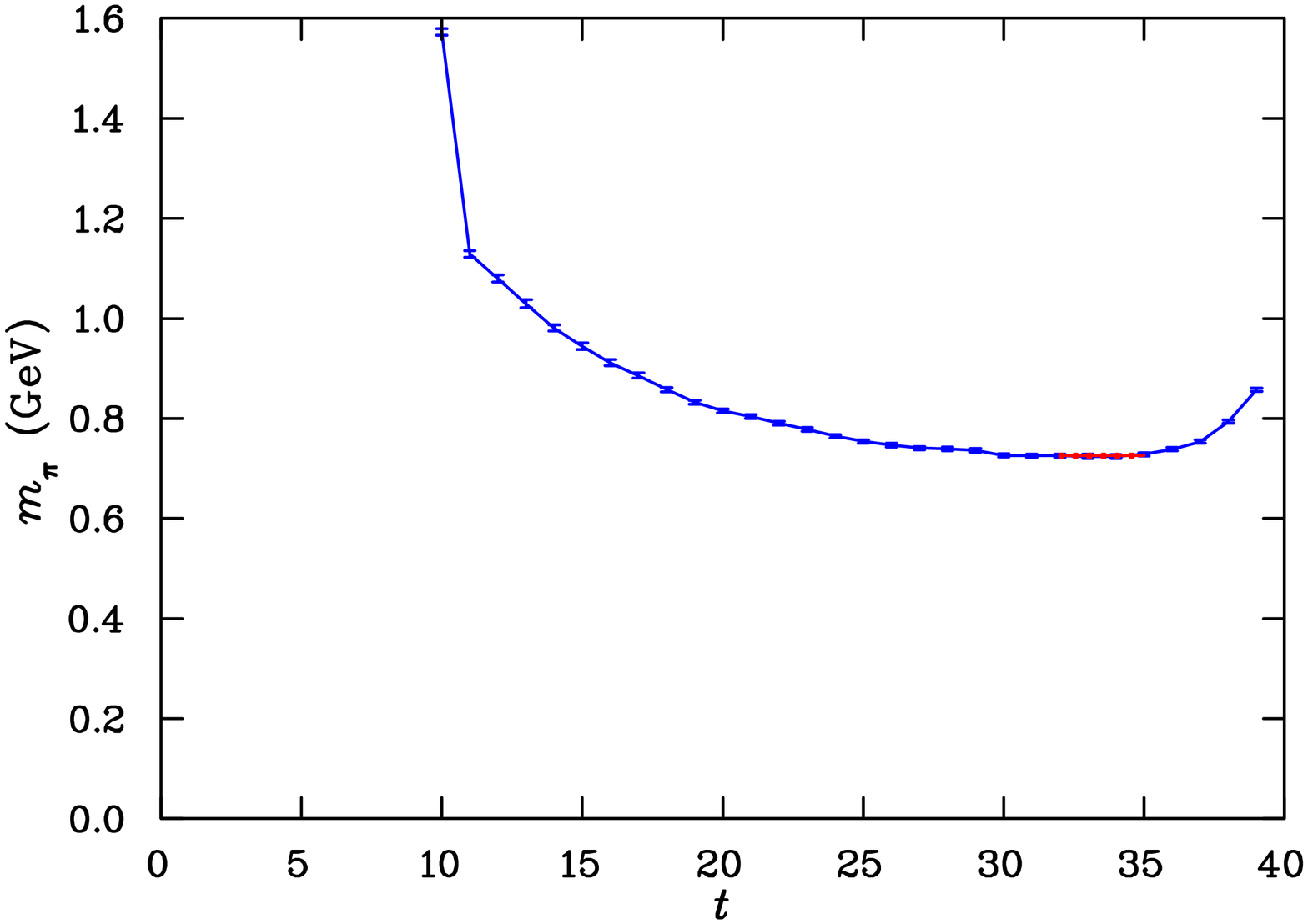}

\includegraphics[height = 0.48\hsize,angle=90]{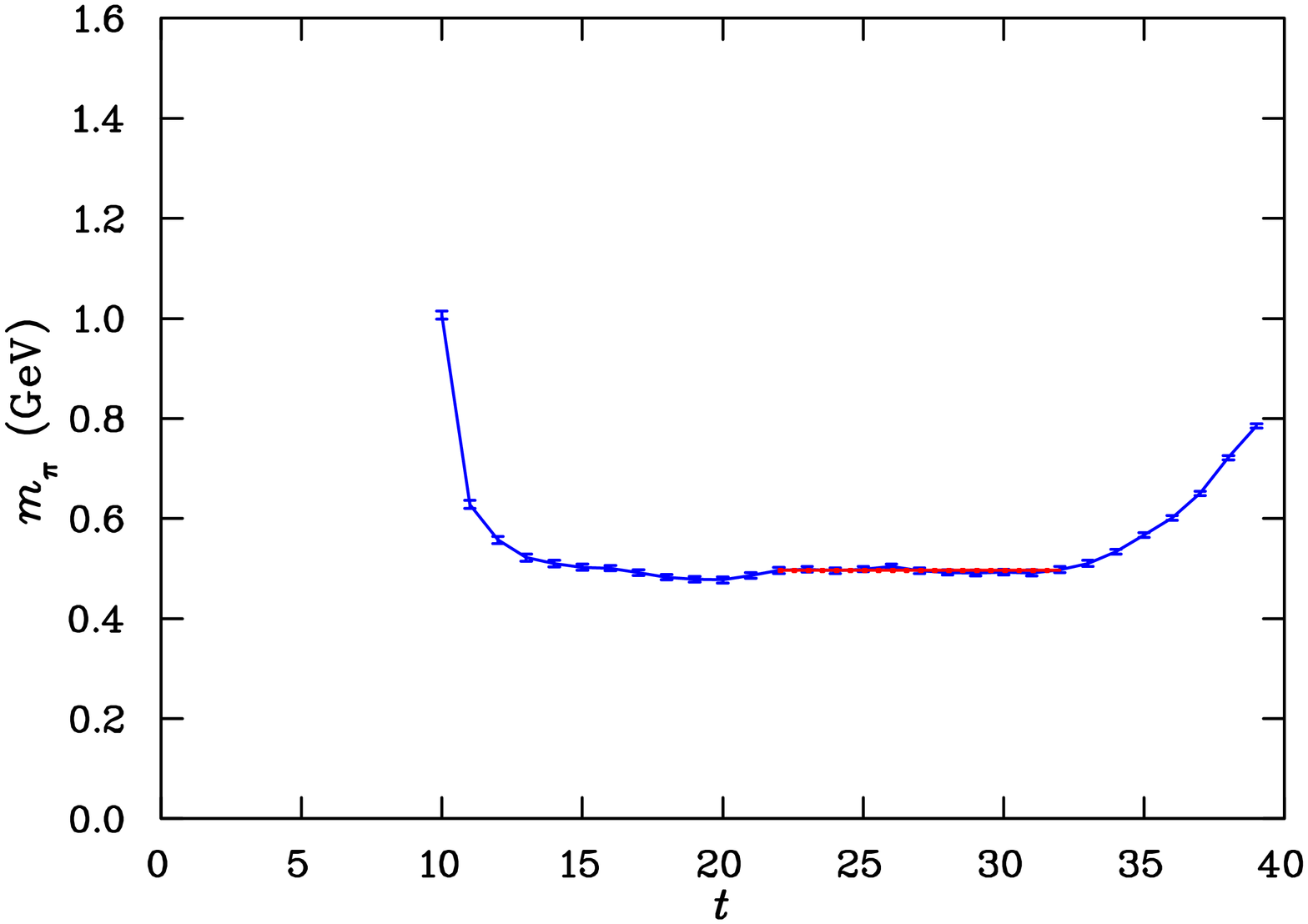}
\hspace{0.09mm}
\includegraphics[height = 0.48\hsize,angle=90]{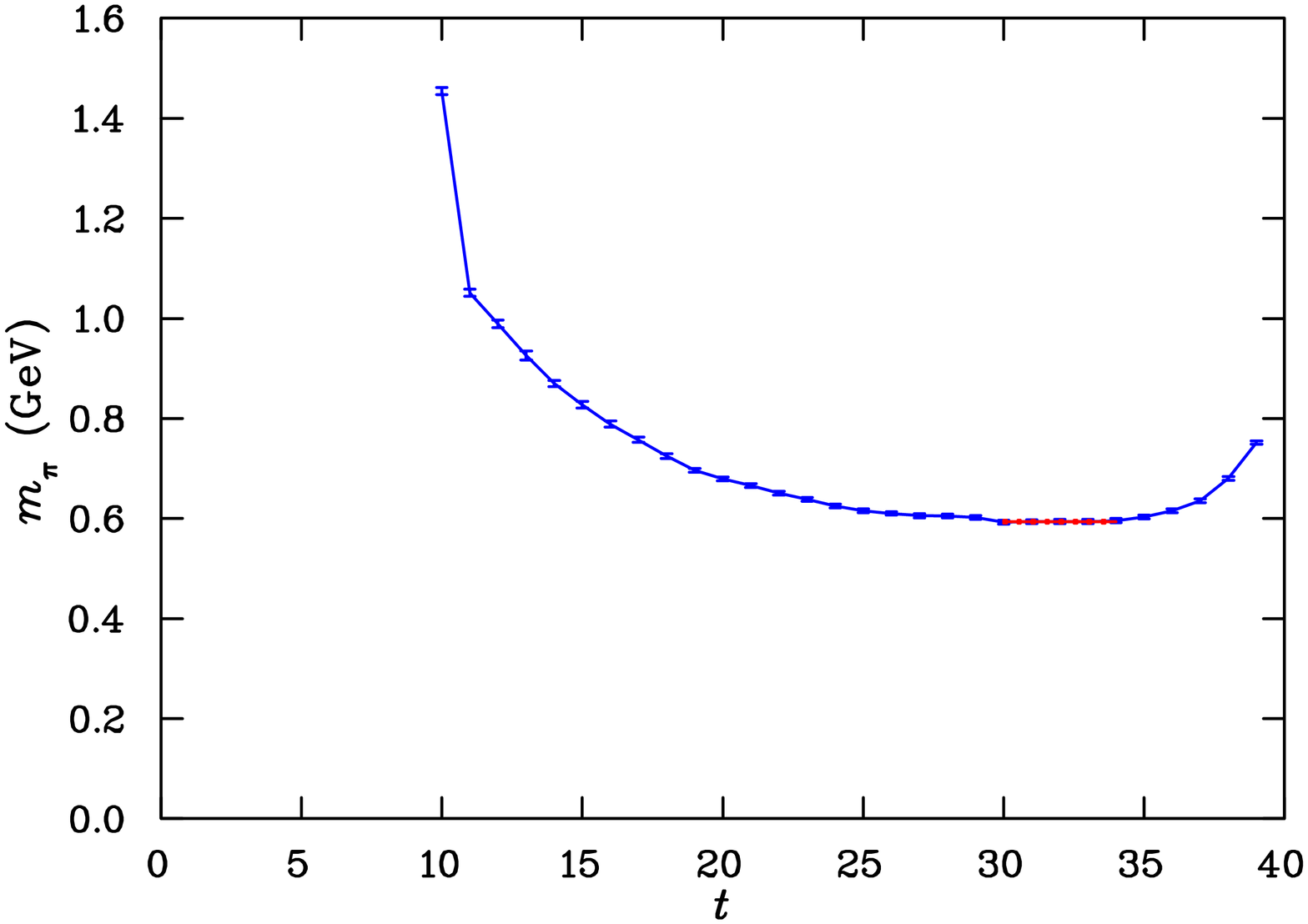}

\includegraphics[height = 0.48\hsize,angle=90]{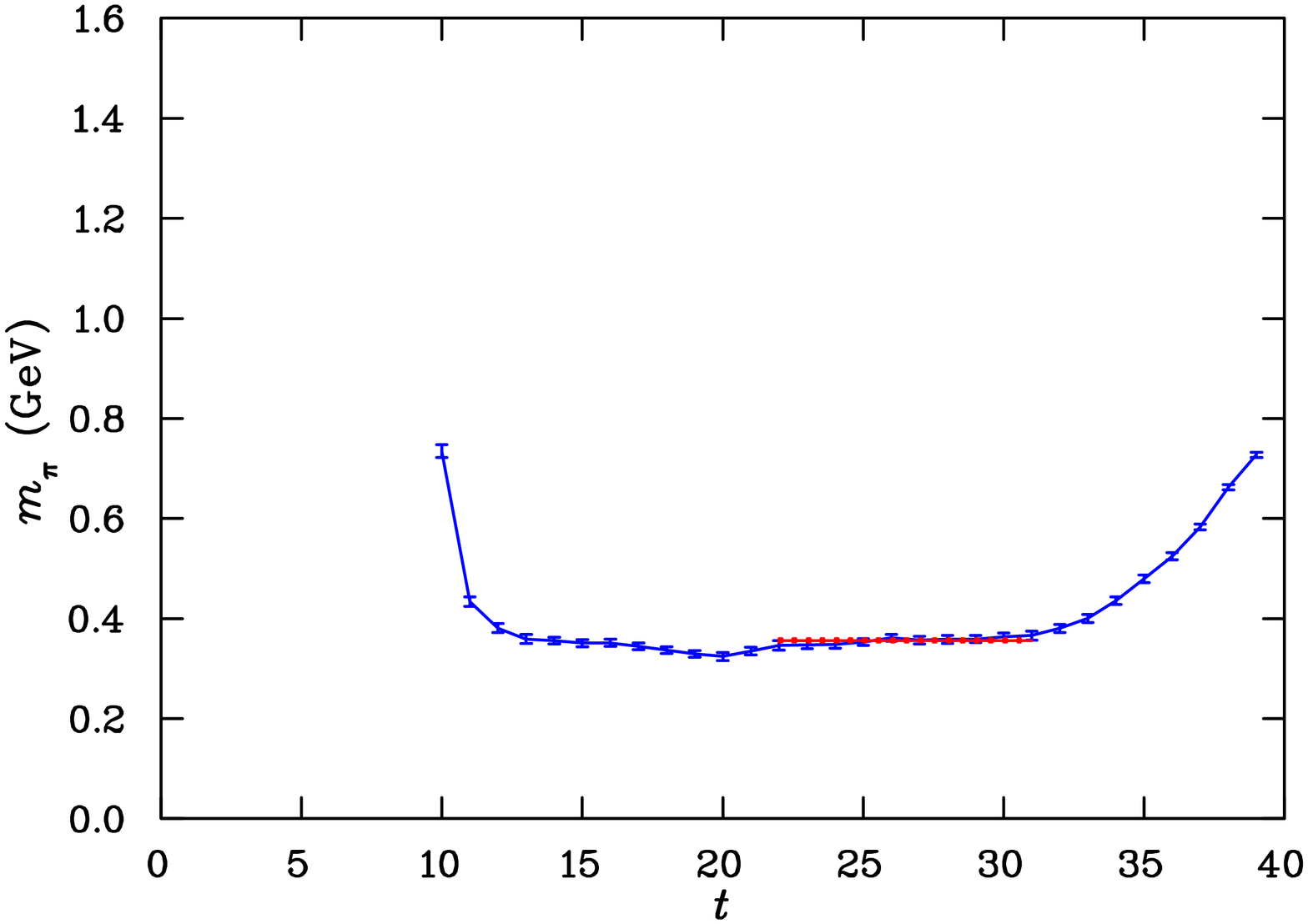}
\hspace{0.09mm}
\includegraphics[height = 0.48\hsize,angle=90]{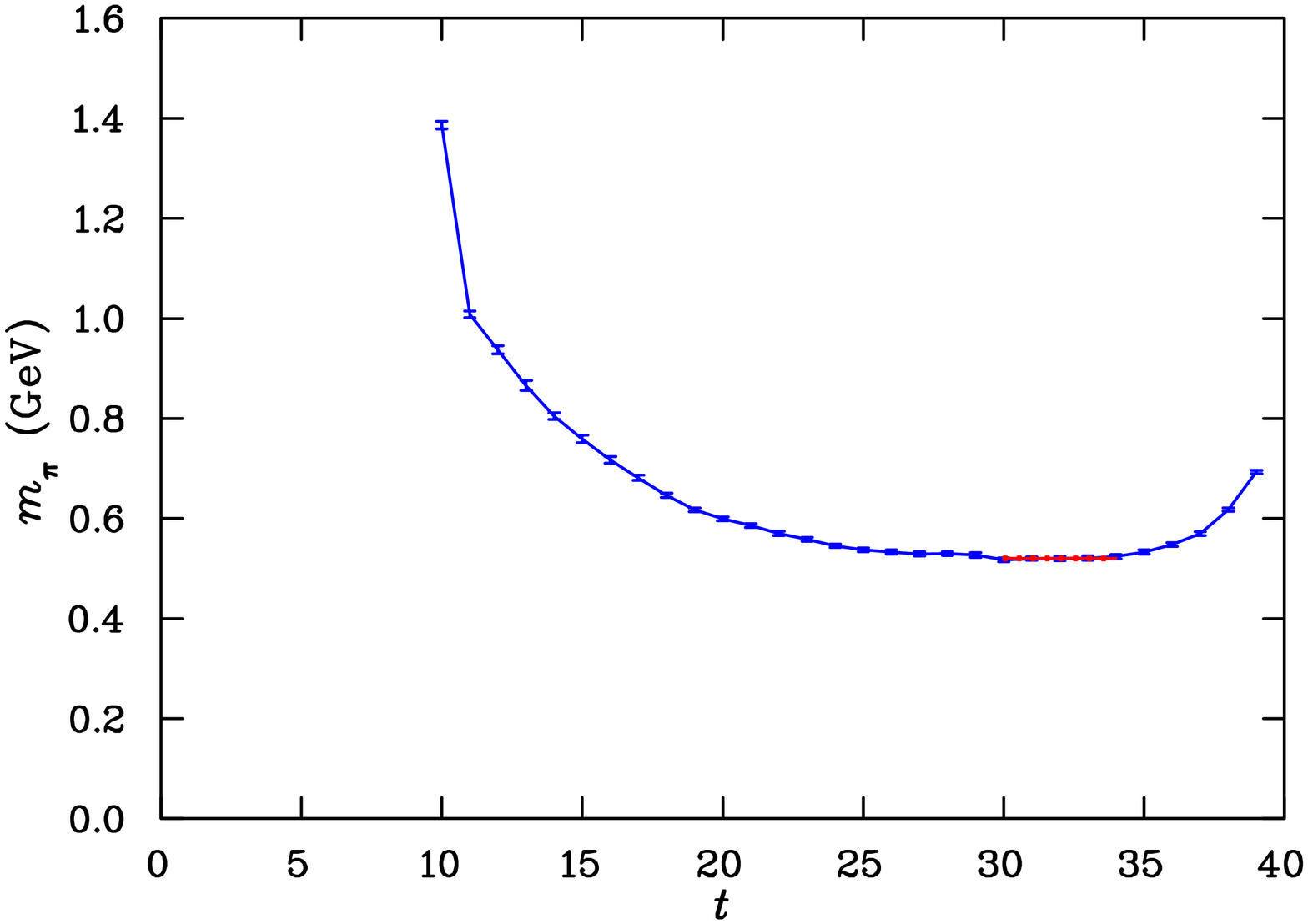}
\caption{Comparison of the pion effective mass evolution $m(t)$ for
  the original configurations (left) and the vortex-free
  configurations (right) as the quark mass is decreased from the top downwards. 
  The values of the hopping parameter $\kappa$ are 0.1280 (top), 0.1293, 0.1304 and
  0.1310 (bottom).}
\label{PionMass}
\end{figure}

We first examine the pion effective mass as a function of Euclidean
time.  The mass for each $\kappa$ is plotted in figure
\ref{PionMass}, where the left column shows the normal configurations
and the right shows the vortex-free configurations.

\begin{figure}[tbh]
\centering
\includegraphics[height = 0.48\hsize,angle=90]{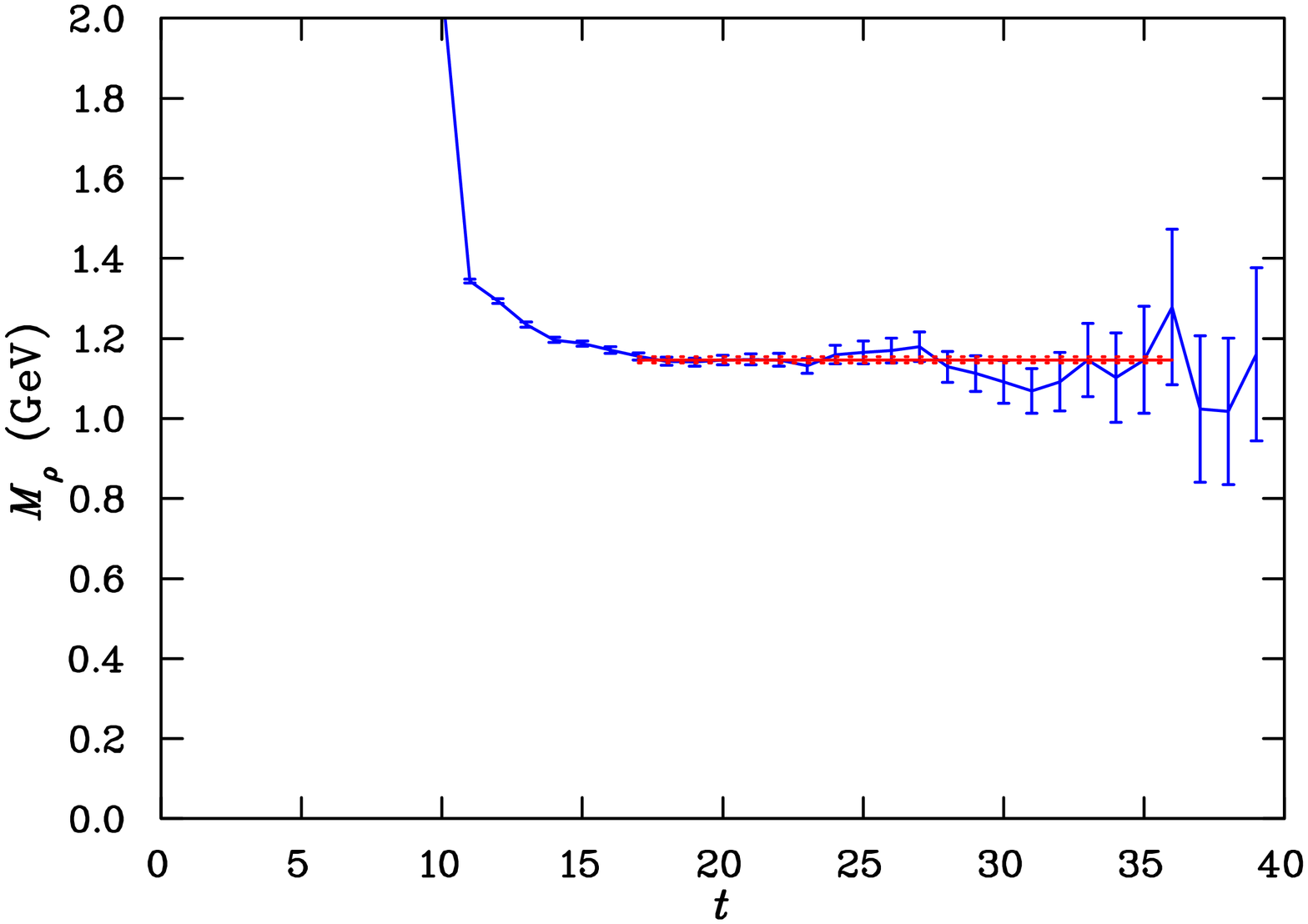} 
\hspace{0.09mm}
\includegraphics[height = 0.48\hsize,angle=90]{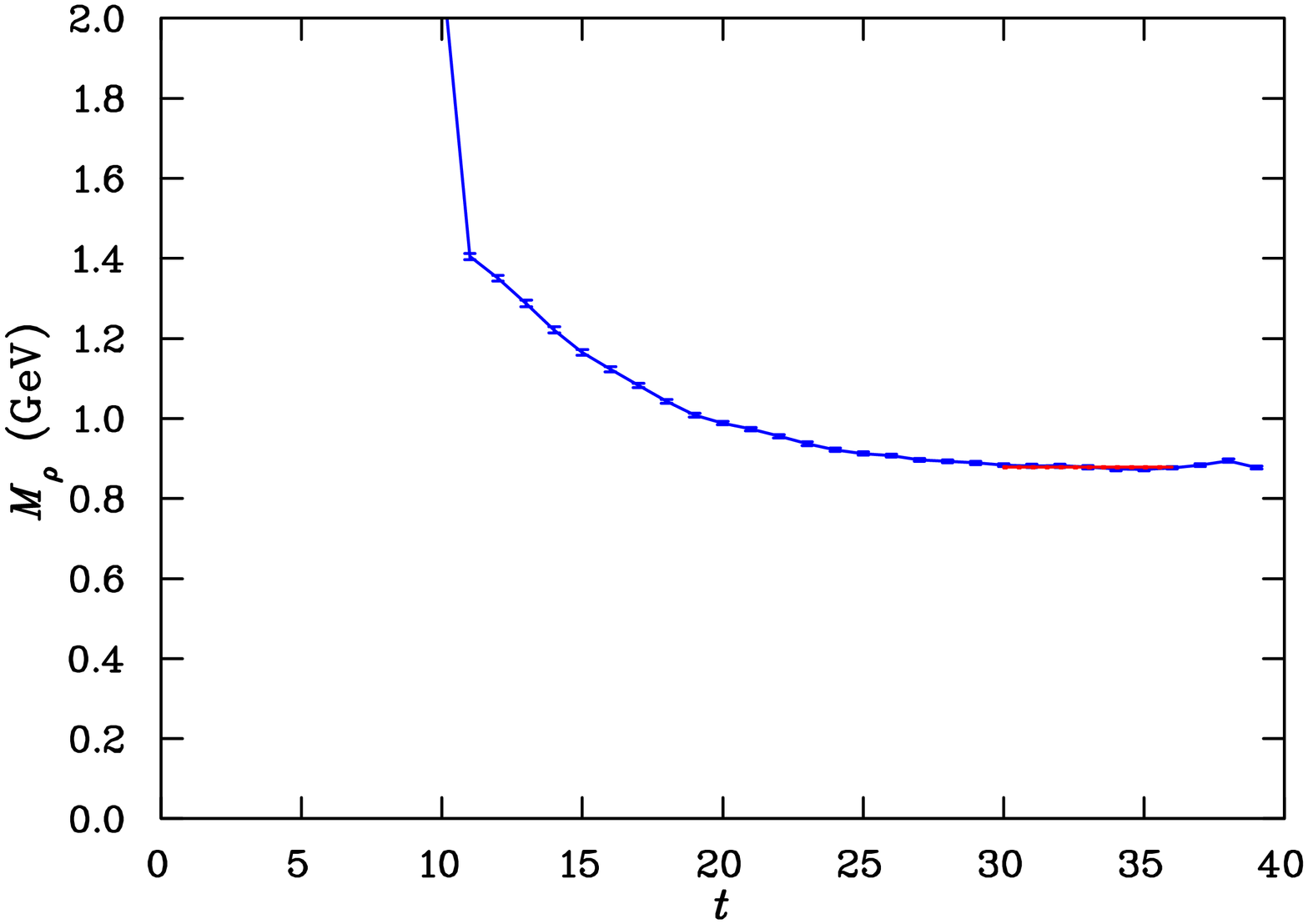}

\includegraphics[height = 0.48\hsize,angle=90]{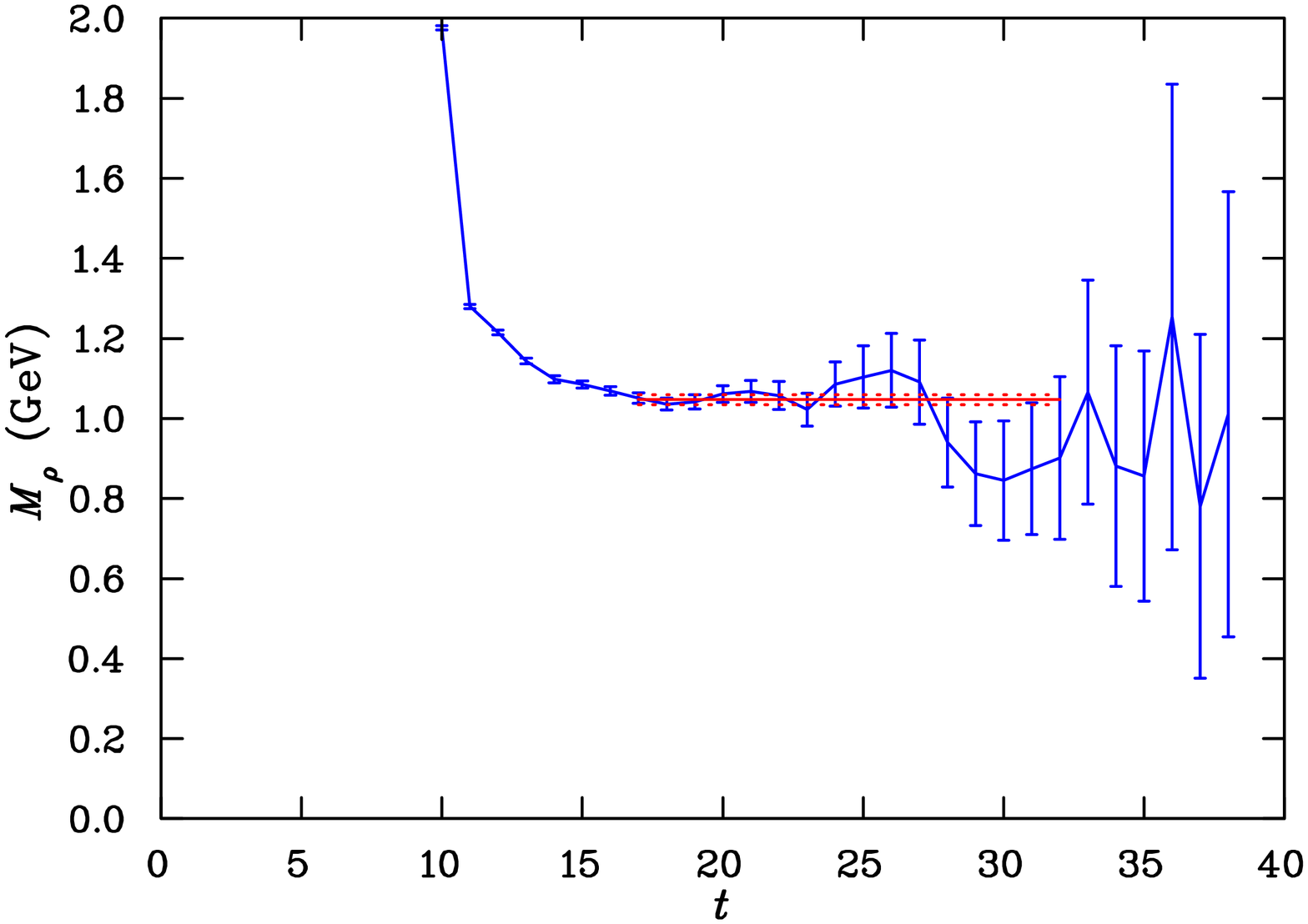} 
\hspace{0.09mm}
\includegraphics[height = 0.48\hsize,angle=90]{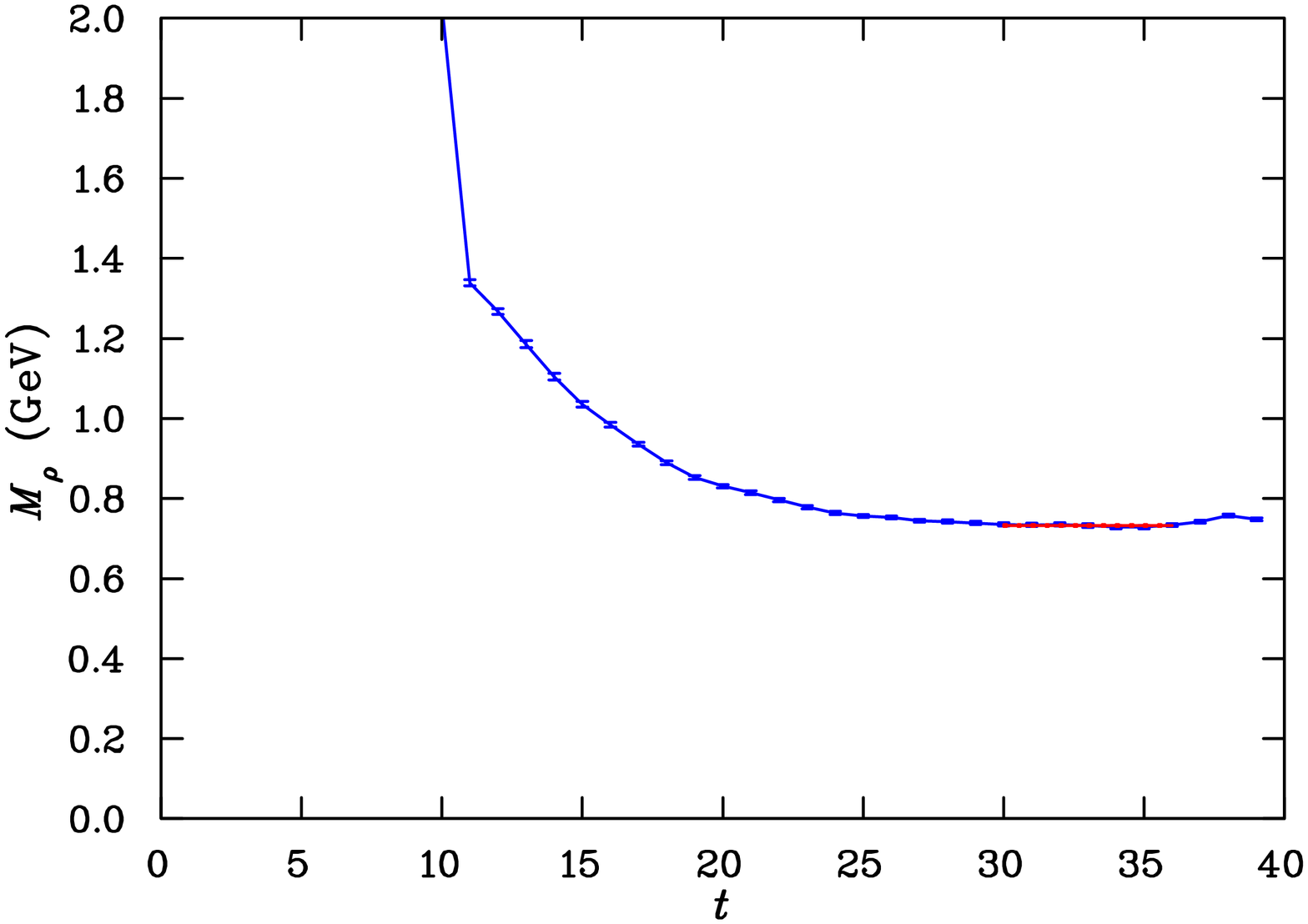}
\caption{Comparison of the $\rho$-meson effective mass evolution
  $m(t)$ for the original configurations (left) and the vortex-free
  configurations (right) as the quark mass is decreased from the top downwards.
  The values of the hopping parameter $\kappa$ are 0.1280 (top) and  0.1293 (bottom).}
\label{RhoMass}
\end{figure}

The pion results reveal an important difference between the two sets
of configurations in the approach to the mass plateau.  The plateau is
approached rapidly on the original configurations, indicating the
presence of a significant mass gap between the ground state and the
first excited state excited by the standard pseudoscalar interpolating
field.  In contrast the approach to the plateau on the vortex-free
configurations is slow, suggesting a tower of closely spaced pion
excitations.  Indeed the shape is reminiscent of the free-field
two-point correlator where excitations are associated with free quarks
having back-to-back momenta of increasing values.  These phenomena are
also observed for the $\rho$-meson effective-mass evolution presented
in Fig.~\ref{RhoMass}.

In the case of the pion, the effective masses from the vortex-free
configurations sit much higher than the regular masses; they do not
show the same low mass associated with the pseudo Goldstone boson of
QCD.

\begin{figure}[tbh]
\centering
\includegraphics[height=\hsize,angle=90]{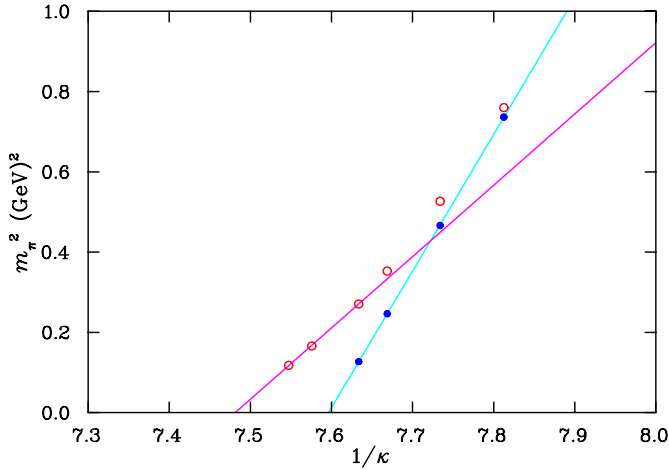}
\caption{Pion mass squared in terms of the inverse hopping
  parameter, $\kappa^{-1}$.  The lines illustrate fits to the original
  configurations and the vortex-free configurations, the latter
  addressing only the lightest three quark masses considered where
  there is some promise that $m_\pi^2 \propto \kappa^{-1}$.}
\label{CompareKappa}
\end{figure}

An examination of the pion masses of the vortex-removed configurations
as a function of the inverse hopping parameter in
Fig.~\ref{CompareKappa} reveals that it is possible to perform
simulations at hopping parameters smaller than the $\kappa_{\rm cr}$
obtained from the original configurations.  This is in accord with
Ref.~\cite{Bowman:2010zr}, where the mass function for the
vortex-removed configurations is shifted higher by about 60 MeV
indicating smaller bare quark masses are required to obtain the same
renormalised quark mass.  We consider two additional hopping parameter
values for the vortex-free configurations which are unphysical for the
normal configurations.  We take $\kappa = 0.1320$, and 0.1325.

This necessarily leads to a different $\kappa_{\rm cr}$
for the vortex-free configurations when using Wilson-style fermions.
Taking the lightest three masses and assuming $m_\pi^2 \propto
1/\kappa$ in the vortex-free theory provides the linear extrapolation
and vortex-free $\kappa_{\rm cr}$ illustrated in
Fig.~\ref{CompareKappa}. Note that the heavier  quark masses in the vortex-free theory show a clear
deviation from linear behaviour.

Of course an alternative scenario is also possible.  One could argue
that dynamical chiral symmetry breaking is spoiled in the vortex-free
theory with $m_\pi^2$ no longer proportional to $1/\kappa$ or $m_q$.
When a quark of mass zero is placed in the vortex-removed
configurations, the pion still has mass.  A comparison of the $\pi$
and $\rho$ meson masses will reveal the correct scenario.

\begin{figure}[tbh]
\centering
\includegraphics[height=\hsize,angle=90]{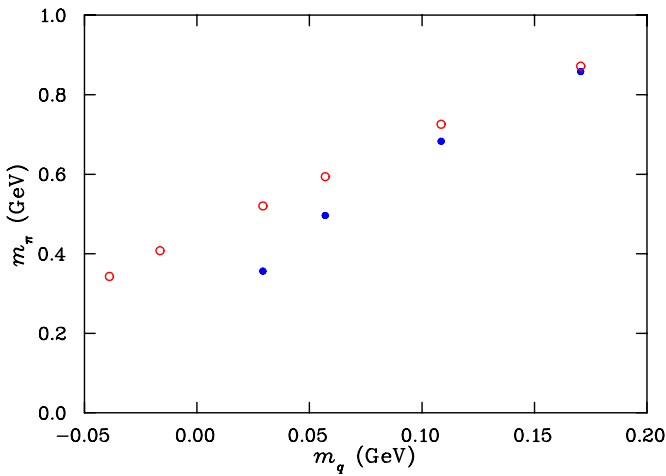}
\caption{Pion mass in GeV as a function of the bare quark mass
  determined with reference to the original configurations.
  Full symbols illustrate results from the original configurations
  while the open symbols illustrate results from the vortex-free
  configurations.  }
\label{Pionmq}
\end{figure}

In Fig.~\ref{Pionmq} the pion mass from the original and vortex-free
configurations is plotted as a function of the bare quark mass, $m_q$
(determined with reference to the critical $\kappa$ value from the
original configurations).  While it seems the vortex-free mass will
approach zero as the quark mass decreases, the relationship between
the quark mass and $m_\pi$ is evidently different between the original
and vortex-free configurations.

The Gell-Mann-Oakes-Renner relationship, $m_\pi^2 \propto m_q$ can be
seen in the results from the original configurations as the points
have the shape of a typical square-root function.  While the pion
masses obtained in the vortex-free configurations at the two lightest
quark masses considered are of a similar magnitude to that of the
lightest pion mass from the original configurations, there is no
evidence of the curvature associated with $m_\pi^2 \propto m_q$.

In the vortex-free configurations the data appears linear with $m_\pi
\propto m_q$ over a wide range of $m_q$, indicating a significant
difference between the two types of configurations and a loss of the
Goldstone nature of the pion in the vortex-free theory.

The ground-state masses for the pion on the regular and vortex-removed
configurations are summarised in Table~\ref{PionMassTable}.

\begin{table}[tbh]
\begin{ruledtabular}
\caption{$\pi$-meson masses from the original and vortex-free
  ensembles as a function of the hopping parameter, $\kappa$.  The
  quark mass, $m_q$ is determined with reference to the original Monte
  Carlo generated configurations. Units are GeV as applicable.\label{PionMassTable}
}
\begin{tabular}{cccc}
         &       &Original   &Vortex-free\\
$\kappa$ & $m_q$ & $m_\pi$ & $m_\pi$ \\
\noalign{\smallskip}
\hline
\noalign{\smallskip}
0.1280 &  0.1705 &0.8582(17)  &0.8717(29) \\
0.1293 &  0.1085 &0.6830(19)  &0.7256(32) \\
0.1304 &  0.0570 &0.4963(24)  &0.5939(28) \\
0.1310 &  0.0293 &0.3564(34)  &0.5204(29) \\
0.1320 & -0.0164 &unphysical  &0.4077(38) \\
0.1325 & -0.0389 &unphysical  &0.3431(43) \\
\end{tabular}
\end{ruledtabular}
\end{table}

\begin{table}[tbh]
\begin{ruledtabular}
\caption{$\rho$-meson masses from the original and vortex-free ensembles as a
  function of the hopping parameter, $\kappa$.  The quark mass, $m_q$
  is determined with reference to the original configurations.  Units
  are GeV as applicable.\label{RhoMassTable}
}
\begin{tabular}{cccc}
         &       &Original   &Vortex-free\\
$\kappa$ & $m_q$ & $m_\rho$ & $m_\rho$ \\
\noalign{\smallskip}
\hline
\noalign{\smallskip}
0.1280 &  0.1705 &1.146(8)    &0.8781(23)   \\
0.1293 &  0.1085 &1.047(12)   &0.7326(23)   \\
0.1304 &  0.0570 &0.982(16)   &0.6058(23)   \\
0.1310 &  0.0293 &0.933(29)   &0.5350(24)   \\
0.1320 & -0.0164 &unphysical  &0.4128(37)   \\
0.1325 & -0.0389 &unphysical  &0.3492(39)   \\
\end{tabular}
\end{ruledtabular}
\end{table}

\begin{figure}[tbh]
\centering
\includegraphics[height=\hsize,angle=90]{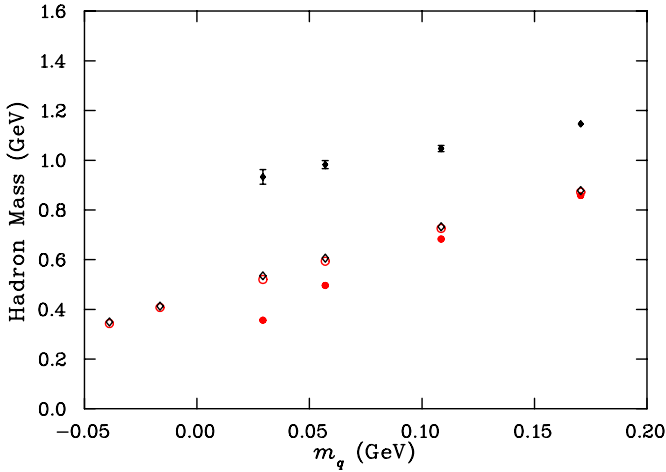}
\caption{Masses for the $\pi$ (circles) and $\rho$ (diamonds) mesons obtained from the
  original configurations (full symbols) and the vortex-free
  configurations (open symbols).  }
\label{PionRhoCompare}
\end{figure}

This loss of a pseudo-Goldstone boson in the vortex-free theory
becomes very clear once one compares the masses of the $\pi$ and
$\rho$ mesons in the vortex-free theory.  Table \ref{RhoMassTable}
reports $\rho$-meson masses and Fig.~\ref{PionRhoCompare} illustrates
masses for the $\pi$ and $\rho$ mesons obtained from the original
configurations (full symbols) and the vortex-free configurations (open
symbols).  This figure clearly illustrates how the $\pi$ and $\rho$
mesons become nearly degenerate on the vortex-free configurations.
Thus the vortex-free pion is not associated with dynamical chiral
symmetry breaking.  It is not a pseudo-Goldstone boson.

\begin{figure}[tbh]
\centering
\includegraphics[height=\hsize,angle=90]{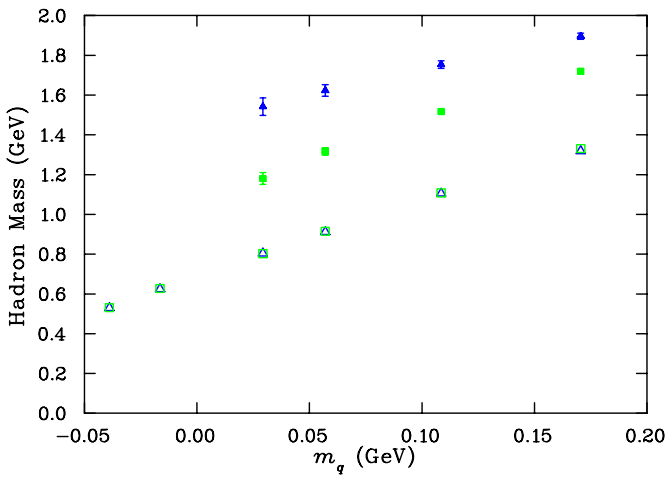}
\caption{Masses of the $\Delta$ (triangles) and $N$ (squares) from the
  original gauge-field configurations are compared with the masses
  from the vortex-free configurations.  Full symbols
  illustrate results from the original configurations while the open
  symbols illustrate results from the vortex-free configurations.  }
\label{DeltaNuc}
\end{figure}

The degeneracy of the $\pi$ and $\rho$ meson is somewhat surprising.
For example, in a simple quark model the $\rho$-meson mass sits much
higher than that of the pion due to a large hyperfine interaction
between the quark and anti-quark.  The degeneracy of the masses on the
vortex-removed configurations implies that any hyperfine interactions
have also been removed with the removal of the centre vortices.

Turning our attention to the baryon sector, the nucleon and $\Delta$
are presented in Fig.~\ref{DeltaNuc} for both the original and the
vortex-free configurations.  From this graph we see that the
vortex-free masses for the $N$ and $\Delta$ are approximately
degenerate, similar to the case of the $\rho$ and $\pi$ mesons.  Again
we see the absence of hyperfine interactions in the vortex-free
theory.  Moreover, both baryons have much lower masses in the
vortex-free theory.  Numerical values for the $N$ and $\Delta$ are
provided in Tables \ref{NucleonMassTable} and \ref{DeltaMassTable}
respectively. 

\begin{table}[tbh]
\begin{ruledtabular}
\caption{Nucleon masses from the original and vortex-free ensembles as a
  function of the hopping parameter, $\kappa$.  The quark mass, $m_q$
  is determined with reference to the original configurations.  Units
  are GeV as applicable.\label{NucleonMassTable}
}
\begin{tabular}{cccc}
         &       &Original   &Vortex-free\\
$\kappa$ & $m_q$ & $m_N$ & $m_N$ \\
\noalign{\smallskip}
\hline
\noalign{\smallskip}
0.1280 &  0.1705 &1.720(12)   &1.3309(43)   \\
0.1293 &  0.1085 &1.517(14)   &1.1078(47)   \\
0.1304 &  0.0570 &1.317(19)   &0.9150(47)   \\
0.1310 &  0.0293 &1.181(30)   &0.8026(48)   \\
0.1320 & -0.0164 &unphysical  &0.6275(53)   \\
0.1325 & -0.0389 &unphysical  &0.5309(56)   \\
\end{tabular}
\end{ruledtabular}
\end{table}

\begin{table}[tbh]
\begin{ruledtabular}
\caption{$\Delta$ baryon masses from the original and vortex-free
  ensembles as a function of the hopping parameter, $\kappa$.  The
  quark mass, $m_q$ is determined with reference to the original
  configurations.  Units are GeV as applicable.\label{DeltaMassTable}
}
\begin{tabular}{cccc}
         &       &Original   &Vortex-free\\
$\kappa$ & $m_q$ & $m_\Delta$ & $m_\Delta$ \\
\noalign{\smallskip}
\hline
\noalign{\smallskip}
0.1280 &  0.1705 &1.896(16)   &1.3187(48)   \\
0.1293 &  0.1085 &1.753(19)   &1.1054(38)   \\
0.1304 &  0.0570 &1.623(29)   &0.9118(39)   \\
0.1310 &  0.0293 &1.542(44)   &0.8043(39)   \\
0.1320 & -0.0164 &unphysical  &0.6251(45)   \\
0.1325 & -0.0389 &unphysical  &0.5298(46)   \\
\end{tabular}
\end{ruledtabular}
\end{table}

\begin{figure}[tbh]
\centering
\includegraphics[height=\hsize,angle=90]{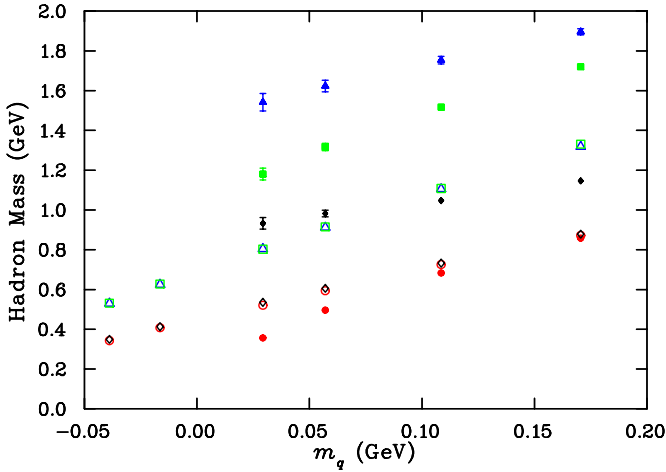}
\caption{Low-lying hadron mass spectrum obtained from the original
  configurations (full symbols) and the vortex-free configurations
  (open symbols).  The symbols correspond to the various hadrons
  considered.  For the original configurations, from lowest to highest
  hadron masses, the symbols correspond to $\pi$, $\rho$, $N$ and
  $\Delta$.  }
\label{All}
\end{figure}

To view the entire hadron mass spectrum we have investigated, all the
hadron masses are plotted on the same axes in Fig.~\ref{All}.  
While hadron masses have become degenerate within the meson and baryon
sectors, it is important to note that significant dynamical mass
generation is occurring.  The dynamical mass generation observed in
the quark mass function of the nonperturbative quark propagator is
also manifest here.

Consider for example, hadron masses at the heaviest quark mass, as
this value has the most accuracy.  Here, the input quark mass is 0.17
GeV.  In a free theory, the meson mass would be 0.34 GeV and the
baryon mass would be 0.51 GeV.  These values are much less than the
measured masses of 0.87 GeV and 1.32 GeV for the meson and baryon sectors
respectively, indicating that whilst the particles have become
degenerate, dynamical mass generation is still present.

The mass generation is reminiscent of the early constituent-quark
model where current quarks are thought to be dressed by QCD-vacuum
interactions giving rise to a constituent quark mass.  This is done in
a model that does not have a connection to chiral symmetry.

\begin{figure}[tbh]
\centering
\includegraphics[height=\hsize,angle=90]{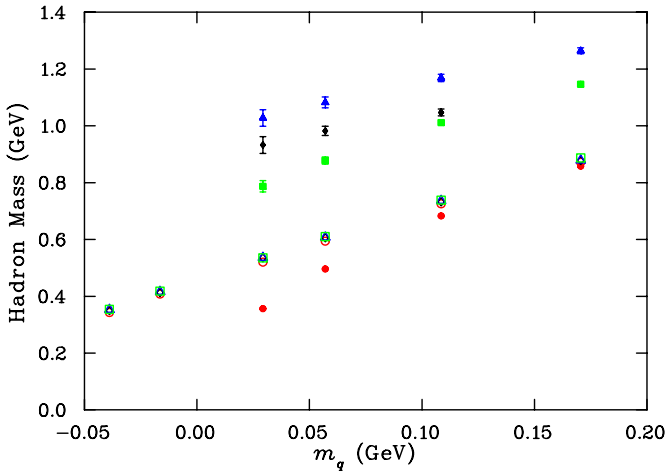}
\caption{The low-lying meson mass spectrum is compared with the
  low-lying baryon mass spectrum multiplied by 2/3 to account for the
  number of constituent quarks composing the system.  Results from the
  original configurations (full symbols) and the vortex-free
  configurations (open symbols) are illustrated.  The symbols are as
  in Fig.~\ref{All}.  }
\label{scaled}
\end{figure}

The apparent degeneracy of the masses from the vortex-free
configurations indicates perhaps that the hadron mass being measured
is merely the sum of the dressed constituent-quark-like masses of the
quarks composing the hadron.  Taking into account the number of
constituent quarks composing each hadron, Fig.~\ref{scaled}
illustrates the masses of the pion, rho meson, 2/3 of the nucleon mass
and 2/3 of the Delta mass as a function of quark mass.  This graph
reveals all four hadrons having the same mass in the vortex-free
theory after one accounts for the number of quarks required to compose
the quantum numbers of the state.  The vortex-free theory is simply a
theory of weakly interacting constituent quarks.

The absence of interactions to lift the degeneracy of the states is
somewhat unexpected.  While the static quark potential loses
confinement in the vortex-free theory, the short distance,
Coulomb-like interactions persist.  Indeed it is the associated
one-gluon-exchange interactions that motivated the spin-dependent
interactions such as spin-orbit or spin-spin hyperfine interactions of
early quark models.  And while these hadron masses have been
calculated in an environment which retains the Coulombic part of the
potential there is no evidence of the associated spin-dependent
interactions.  Either, the short-distance noise introduced in the
process of centre-vortex removal has spoiled these short-distance
interactions or the dominant origin of spin-dependent
interactions is in the confining part of the potential.

\section{Conclusions}

This study resolves an important question on the role of centre
vortices in dynamical chiral symmetry breaking in $SU(3)$ gauge theory.
Is the persistence of dynamical mass generation in the mass function
of the quark propagator a manifestation of dynamical chiral symmetry
breaking in the absence of confinement?

Prior to this work, studies of the quark propagator in $SU(3)$ gauge
theory \cite{Bowman:2010zr} admitted the possibility that the
underlying mechanisms generating confinement and dynamical chiral
symmetry breaking were decoupled as dynamical mass generation survives
in the absence of confinement.  This work resolves this issue by
revealing that the dynamical mass generation observed in the
vortex-free theory is not associated with chiral symmetry.

A comparison of the input quark mass and the hadron masses reveals
that dynamical mass generation is at work.  This is in accord with
Ref.~\cite{Bowman:2010zr} which clearly illustrates how dynamical-mass
generation survives the removal of centre vortices.

However, of greatest importance is the complete absence of any remnant
of dynamical chiral symmetry breaking in the mass spectrum of the
vortex-free theory.  We find a pion degenerate with the $\rho$ meson
and a mass dependence of $m_\pi \propto m_q$ inconsistent with the
properties of the pseudo-Goldstone boson of chiral symmetry.  Because
these key results are drawn from the vortex-free theory alone the
results are robust, independent of any uncertainty in defining the
critical hopping parameter in the vortex-free theory.

From our calculations of the hadron mass spectrum, we have seen that
the hadron masses of the vortex-free theory are simply a reflection of
the number of quarks required to compose their quantum numbers.  There
is little evidence of quark interactions in the mass spectrum and this
is in accord with the general features of the Euclidean time evolution
of the hadron effective masses in the vortex-free theory where the
spectrum suggests a theory of free constituent quarks.

One interesting question which remains is the nature of the
vortex-free hadron spectrum at vanishing quark mass.  To explore this
question one must adopt a fermion action for which the chiral limit is
well defined at $m_q = 0$.  For example, both staggered and overlap
fermion formalisms provide this property and it would be interesting
to further examine the vortex-free spectrum with these actions.

In conclusion, centre vortex removal spoils both confinement and chiral
symmetry.  Centre-vortices are the most fundamental degrees of freedom
in QCD, essential to confinement and dynamical chiral symmetry
breaking.  Just as in SU(2), there is an intimate relationship between
centre vortices, confinement and dynamical chiral symmetry breaking.
Both confinement and dynamical chiral symmetry breaking are lost under
centre vortex removal.

\acknowledgments

It is a pleasure to acknowledge the contributions of Kurt Langfeld and
Alan \'O~Cais in centre gauge fixing and centre projecting the gauge
field configurations \cite{Cais:2008za} used in this study.  This
research was undertaken on the NCI National Facility in Canberra,
Australia, which is supported by the Australian Commonwealth
Government. We also acknowledge eResearch SA for grants of
supercomputing time.  This research is supported by the Australian
Research Council.

\bibliography{SU3massSpec}

\begin{thebibliography}{22}
\expandafter\ifx\csname natexlab\endcsname\relax\def\natexlab#1{#1}\fi
\expandafter\ifx\csname bibnamefont\endcsname\relax
  \def\bibnamefont#1{#1}\fi
\expandafter\ifx\csname bibfnamefont\endcsname\relax
  \def\bibfnamefont#1{#1}\fi
\expandafter\ifx\csname citenamefont\endcsname\relax
  \def\citenamefont#1{#1}\fi
\expandafter\ifx\csname url\endcsname\relax
  \def\url#1{\texttt{#1}}\fi
\expandafter\ifx\csname urlprefix\endcsname\relax\def\urlprefix{URL }\fi
\providecommand{\bibinfo}[2]{#2}
\providecommand{\eprint}[2][]{\url{#2}}

\bibitem[{\citenamefont{'t~Hooft}(1976)}]{'tHooft:1976fv}
\bibinfo{author}{\bibfnamefont{G.}~\bibnamefont{'t~Hooft}},
  \bibinfo{journal}{Phys.Rev.} \textbf{\bibinfo{volume}{D14}},
  \bibinfo{pages}{3432} (\bibinfo{year}{1976}).

\bibitem[{\citenamefont{Banks and Casher}(1980)}]{Banks:1979yr}
\bibinfo{author}{\bibfnamefont{T.}~\bibnamefont{Banks}} \bibnamefont{and}
  \bibinfo{author}{\bibfnamefont{A.}~\bibnamefont{Casher}},
  \bibinfo{journal}{Nucl.Phys.} \textbf{\bibinfo{volume}{B169}},
  \bibinfo{pages}{103} (\bibinfo{year}{1980}), \bibinfo{note}{revised Version}.

\bibitem[{\citenamefont{Chen et~al.}(1999)\citenamefont{Chen, Brower, Negele,
  and Shuryak}}]{Chen:1998ct}
\bibinfo{author}{\bibfnamefont{D.}~\bibnamefont{Chen}},
  \bibinfo{author}{\bibfnamefont{R.}~\bibnamefont{Brower}},
  \bibinfo{author}{\bibfnamefont{J.~W.} \bibnamefont{Negele}},
  \bibnamefont{and} \bibinfo{author}{\bibfnamefont{E.~V.}
  \bibnamefont{Shuryak}}, \bibinfo{journal}{Nucl.Phys.Proc.Suppl.}
  \textbf{\bibinfo{volume}{73}}, \bibinfo{pages}{512} (\bibinfo{year}{1999}),
  \eprint{hep-lat/9809091}.

\bibitem[{\citenamefont{'t~Hooft}(1981)}]{'tHooft:1981ht}
\bibinfo{author}{\bibfnamefont{G.}~\bibnamefont{'t~Hooft}},
  \bibinfo{journal}{Nucl. Phys.} \textbf{\bibinfo{volume}{B190}},
  \bibinfo{pages}{455} (\bibinfo{year}{1981}).

\bibitem[{\citenamefont{Mandelstam}(1976)}]{Mandelstam:1974pi}
\bibinfo{author}{\bibfnamefont{S.}~\bibnamefont{Mandelstam}},
  \bibinfo{journal}{Phys.Rept.} \textbf{\bibinfo{volume}{23}},
  \bibinfo{pages}{245} (\bibinfo{year}{1976}).

\bibitem[{\citenamefont{Shiba and Suzuki}(1994)}]{Shiba:1994ab}
\bibinfo{author}{\bibfnamefont{H.}~\bibnamefont{Shiba}} \bibnamefont{and}
  \bibinfo{author}{\bibfnamefont{T.}~\bibnamefont{Suzuki}},
  \bibinfo{journal}{Phys.Lett.} \textbf{\bibinfo{volume}{B333}},
  \bibinfo{pages}{461} (\bibinfo{year}{1994}), \eprint{hep-lat/9404015}.

\bibitem[{\citenamefont{de~Forcrand and D'Elia}(1999)}]{deForcrand:1999ms}
\bibinfo{author}{\bibfnamefont{P.}~\bibnamefont{de~Forcrand}} \bibnamefont{and}
  \bibinfo{author}{\bibfnamefont{M.}~\bibnamefont{D'Elia}},
  \bibinfo{journal}{Phys. Rev. Lett.} \textbf{\bibinfo{volume}{82}},
  \bibinfo{pages}{4582} (\bibinfo{year}{1999}), \eprint{hep-lat/9901020}.

\bibitem[{\citenamefont{Bowman et~al.}(2008)\citenamefont{Bowman, Langfeld,
  Leinweber, O'~Cais, Sternbeck et~al.}}]{Bowman:2008qd}
\bibinfo{author}{\bibfnamefont{P.~O.} \bibnamefont{Bowman}},
  \bibinfo{author}{\bibfnamefont{K.}~\bibnamefont{Langfeld}},
  \bibinfo{author}{\bibfnamefont{D.~B.} \bibnamefont{Leinweber}},
  \bibinfo{author}{\bibfnamefont{A.}~\bibnamefont{O'~Cais}},
  \bibinfo{author}{\bibfnamefont{A.}~\bibnamefont{Sternbeck}},
  \bibnamefont{et~al.}, \bibinfo{journal}{Phys.Rev.}
  \textbf{\bibinfo{volume}{D78}}, \bibinfo{pages}{054509}
  (\bibinfo{year}{2008}), \eprint{0806.4219}.

\bibitem[{\citenamefont{Leinweber et~al.}(2006)\citenamefont{Leinweber, Bowman,
  Heller, Kusterer, Langfeld et~al.}}]{Leinweber:2006zq}
\bibinfo{author}{\bibfnamefont{D.}~\bibnamefont{Leinweber}},
  \bibinfo{author}{\bibfnamefont{P.}~\bibnamefont{Bowman}},
  \bibinfo{author}{\bibfnamefont{U.}~\bibnamefont{Heller}},
  \bibinfo{author}{\bibfnamefont{D.}~\bibnamefont{Kusterer}},
  \bibinfo{author}{\bibfnamefont{K.}~\bibnamefont{Langfeld}},
  \bibnamefont{et~al.}, \bibinfo{journal}{Nucl.Phys.Proc.Suppl.}
  \textbf{\bibinfo{volume}{161}}, \bibinfo{pages}{130} (\bibinfo{year}{2006}).

\bibitem[{\citenamefont{Cais et~al.}(2007)\citenamefont{Cais, Kamleh, Lasscock,
  Leinweber, von Smekal et~al.}}]{Cais:2007bm}
\bibinfo{author}{\bibfnamefont{A.~O.} \bibnamefont{Cais}},
  \bibinfo{author}{\bibfnamefont{W.}~\bibnamefont{Kamleh}},
  \bibinfo{author}{\bibfnamefont{B.}~\bibnamefont{Lasscock}},
  \bibinfo{author}{\bibfnamefont{D.}~\bibnamefont{Leinweber}},
  \bibinfo{author}{\bibfnamefont{L.}~\bibnamefont{von Smekal}},
  \bibnamefont{et~al.}, \bibinfo{journal}{PoS}
  \textbf{\bibinfo{volume}{LAT2007}}, \bibinfo{pages}{321}
  (\bibinfo{year}{2007}), \eprint{0710.2958}.

\bibitem[{\citenamefont{Bowman et~al.}(2011)\citenamefont{Bowman, Langfeld,
  Leinweber, Sternbeck, von Smekal et~al.}}]{Bowman:2010zr}
\bibinfo{author}{\bibfnamefont{P.~O.} \bibnamefont{Bowman}},
  \bibinfo{author}{\bibfnamefont{K.}~\bibnamefont{Langfeld}},
  \bibinfo{author}{\bibfnamefont{D.~B.} \bibnamefont{Leinweber}},
  \bibinfo{author}{\bibfnamefont{A.}~\bibnamefont{Sternbeck}},
  \bibinfo{author}{\bibfnamefont{L.}~\bibnamefont{von Smekal}},
  \bibnamefont{et~al.}, \bibinfo{journal}{Phys.Rev.}
  \textbf{\bibinfo{volume}{D84}}, \bibinfo{pages}{034501}
  (\bibinfo{year}{2011}), \eprint{1010.4624}.

\bibitem[{\citenamefont{Del~Debbio et~al.}(1998)\citenamefont{Del~Debbio,
  Faber, Giedt, Greensite, and Olejnik}}]{DelDebbio:1998uu}
\bibinfo{author}{\bibfnamefont{L.}~\bibnamefont{Del~Debbio}},
  \bibinfo{author}{\bibfnamefont{M.}~\bibnamefont{Faber}},
  \bibinfo{author}{\bibfnamefont{J.}~\bibnamefont{Giedt}},
  \bibinfo{author}{\bibfnamefont{J.}~\bibnamefont{Greensite}},
  \bibnamefont{and} \bibinfo{author}{\bibfnamefont{S.}~\bibnamefont{Olejnik}},
  \bibinfo{journal}{Phys.Rev.} \textbf{\bibinfo{volume}{D58}},
  \bibinfo{pages}{094501} (\bibinfo{year}{1998}), \eprint{hep-lat/9801027}.

\bibitem[{\citenamefont{O'Malley et~al.}(2011)\citenamefont{O'Malley, Kamleh,
  Leinweber, and Moran}}]{OMalleyLatt2011}
\bibinfo{author}{\bibfnamefont{E.-A.} \bibnamefont{O'Malley}},
  \bibinfo{author}{\bibfnamefont{W.}~\bibnamefont{Kamleh}},
  \bibinfo{author}{\bibfnamefont{D.}~\bibnamefont{Leinweber}},
  \bibnamefont{and} \bibinfo{author}{\bibfnamefont{P.}~\bibnamefont{Moran}},
  \bibinfo{journal}{PoS} \textbf{\bibinfo{volume}{LATTICE 2011}},
  \bibinfo{pages}{257} (\bibinfo{year}{2011}).

\bibitem[{\citenamefont{Del~Debbio et~al.}(1997)\citenamefont{Del~Debbio,
  Faber, Greensite, and Olejnik}}]{DelDebbio:1996mh}
\bibinfo{author}{\bibfnamefont{L.}~\bibnamefont{Del~Debbio}},
  \bibinfo{author}{\bibfnamefont{M.}~\bibnamefont{Faber}},
  \bibinfo{author}{\bibfnamefont{J.}~\bibnamefont{Greensite}},
  \bibnamefont{and} \bibinfo{author}{\bibfnamefont{S.}~\bibnamefont{Olejnik}},
  \bibinfo{journal}{Phys. Rev.} \textbf{\bibinfo{volume}{D55}},
  \bibinfo{pages}{2298} (\bibinfo{year}{1997}), \eprint{hep-lat/9610005}.

\bibitem[{\citenamefont{Langfeld et~al.}(1998)\citenamefont{Langfeld,
  Reinhardt, and Tennert}}]{Langfeld:1997jx}
\bibinfo{author}{\bibfnamefont{K.}~\bibnamefont{Langfeld}},
  \bibinfo{author}{\bibfnamefont{H.}~\bibnamefont{Reinhardt}},
  \bibnamefont{and} \bibinfo{author}{\bibfnamefont{O.}~\bibnamefont{Tennert}},
  \bibinfo{journal}{Phys. Lett.} \textbf{\bibinfo{volume}{B419}},
  \bibinfo{pages}{317} (\bibinfo{year}{1998}), \eprint{hep-lat/9710068}.

\bibitem[{\citenamefont{Langfeld}(2004)}]{Langfeld:2003ev}
\bibinfo{author}{\bibfnamefont{K.}~\bibnamefont{Langfeld}},
  \bibinfo{journal}{Phys. Rev.} \textbf{\bibinfo{volume}{D69}},
  \bibinfo{pages}{014503} (\bibinfo{year}{2004}), \eprint{hep-lat/0307030}.

\bibitem[{\citenamefont{de~Forcrand and Pepe}(2001)}]{deForcrand:2000pg}
\bibinfo{author}{\bibfnamefont{P.}~\bibnamefont{de~Forcrand}} \bibnamefont{and}
  \bibinfo{author}{\bibfnamefont{M.}~\bibnamefont{Pepe}},
  \bibinfo{journal}{Nucl. Phys.} \textbf{\bibinfo{volume}{B598}},
  \bibinfo{pages}{557} (\bibinfo{year}{2001}), \eprint{hep-lat/0008016}.

\bibitem[{\citenamefont{O'Cais et~al.}(2010)\citenamefont{O'Cais, Kamleh,
  Langfeld, Lasscock, Leinweber et~al.}}]{Cais:2008za}
\bibinfo{author}{\bibfnamefont{A.}~\bibnamefont{O'Cais}},
  \bibinfo{author}{\bibfnamefont{W.}~\bibnamefont{Kamleh}},
  \bibinfo{author}{\bibfnamefont{K.}~\bibnamefont{Langfeld}},
  \bibinfo{author}{\bibfnamefont{B.}~\bibnamefont{Lasscock}},
  \bibinfo{author}{\bibfnamefont{D.}~\bibnamefont{Leinweber}},
  \bibnamefont{et~al.}, \bibinfo{journal}{Phys.Rev.}
  \textbf{\bibinfo{volume}{D82}}, \bibinfo{pages}{114512}
  (\bibinfo{year}{2010}), \eprint{0807.0264}.

\bibitem[{\citenamefont{L{\"u}scher and Weisz}(1985)}]{Luscher:1984xn}
\bibinfo{author}{\bibfnamefont{M.}~\bibnamefont{L{\"u}scher}} \bibnamefont{and}
  \bibinfo{author}{\bibfnamefont{P.}~\bibnamefont{Weisz}},
  \bibinfo{journal}{Commun. Math. Phys.} \textbf{\bibinfo{volume}{97}},
  \bibinfo{pages}{59} (\bibinfo{year}{1985}).

\bibitem[{\citenamefont{Zanotti et~al.}(2002)}]{Zanotti:2001yb}
\bibinfo{author}{\bibfnamefont{J.~M.} \bibnamefont{Zanotti}}
  \bibnamefont{et~al.} (\bibinfo{collaboration}{CSSM Lattice Collaboration}),
  \bibinfo{journal}{Phys.Rev.} \textbf{\bibinfo{volume}{D65}},
  \bibinfo{pages}{074507} (\bibinfo{year}{2002}), \eprint{hep-lat/0110216}.

\bibitem[{\citenamefont{Zanotti et~al.}(2005)\citenamefont{Zanotti, Lasscock,
  Leinweber, and Williams}}]{Zanotti:2004dr}
\bibinfo{author}{\bibfnamefont{J.}~\bibnamefont{Zanotti}},
  \bibinfo{author}{\bibfnamefont{B.}~\bibnamefont{Lasscock}},
  \bibinfo{author}{\bibfnamefont{D.}~\bibnamefont{Leinweber}},
  \bibnamefont{and} \bibinfo{author}{\bibfnamefont{A.}~\bibnamefont{Williams}},
  \bibinfo{journal}{Phys.Rev.} \textbf{\bibinfo{volume}{D71}},
  \bibinfo{pages}{034510} (\bibinfo{year}{2005}), \eprint{hep-lat/0405015}.

\bibitem[{\citenamefont{Boinepalli et~al.}(2005)\citenamefont{Boinepalli,
  Kamleh, Leinweber, Williams, and Zanotti}}]{Boinepalli:2004fz}
\bibinfo{author}{\bibfnamefont{S.}~\bibnamefont{Boinepalli}},
  \bibinfo{author}{\bibfnamefont{W.}~\bibnamefont{Kamleh}},
  \bibinfo{author}{\bibfnamefont{D.~B.} \bibnamefont{Leinweber}},
  \bibinfo{author}{\bibfnamefont{A.~G.} \bibnamefont{Williams}},
  \bibnamefont{and} \bibinfo{author}{\bibfnamefont{J.~M.}
  \bibnamefont{Zanotti}}, \bibinfo{journal}{Phys.Lett.}
  \textbf{\bibinfo{volume}{B616}}, \bibinfo{pages}{196} (\bibinfo{year}{2005}),
  \eprint{hep-lat/0405026}.

\end{thebibliography}

\end{document}